\documentclass[a4paper]{jpconf}

\usepackage{graphicx}
\usepackage{grffile}
\usepackage[%
  sumlimits,
  intlimits,
  namelimits]{amsmath}
\usepackage{mathtools}
\usepackage{amssymb}
\usepackage{amsfonts}
\usepackage{amsxtra}

\makeatletter
\g@addto@macro\bfseries{\boldmath}
\makeatother

\begin{document}

\title{Meson Spectroscopy at COMPASS}

\author{Boris Grube$^1$ on behalf of the COMPASS collaboration}

\address{$^1$ Physik-Department E18, Technische Universit\"at M\"unchen, Garching, Germany}

\ead{bgrube@tum.de}

\begin{abstract}
  The COmmon Muon and Proton Apparatus for Structure and Spectroscopy
  (COMPASS) is a multi-purpose fixed-target experiment at the CERN
  Super Proton Synchrotron (SPS) aimed at studying the structure and
  spectrum of hadrons.  The two-stage spectrometer has a good
  acceptance for charged as well as neutral particles over a wide
  kinematic range and thus allows to access a wide range of reactions.
  Light mesons are studied with negative (mostly $\pi^-$) and positive
  ($p$, $\pi^+$) hadron beams with a momentum of 190~GeV/$c$.  The
  spectrum of light mesons is investigated in various final states
  produced in diffractive dissociation reactions at squared
  four-momentum transfers to the target between 0.1~and
  1.0~$(\text{GeV}/c)^2$.  The flagship channel is the
  $\pi^-\pi^+\pi^-$ final state, for which COMPASS has recorded the
  currently largest data sample.  These data not only allow to measure
  the properties of known resonances with high precision, but also to
  search for new states.  Among these is a new resonance-like signal,
  the $a_1(1420)$, with unusual properties.  Of particular interest is
  also the resonance content of the partial wave with spin-exotic
  $J^{PC} = 1^{-+}$ quantum numbers, which are forbidden for
  quark-antiquark states.
\end{abstract}

\section{Introduction}

\begin{figure}[tb]
  \centering
  \includegraphics[width=0.45\textwidth]{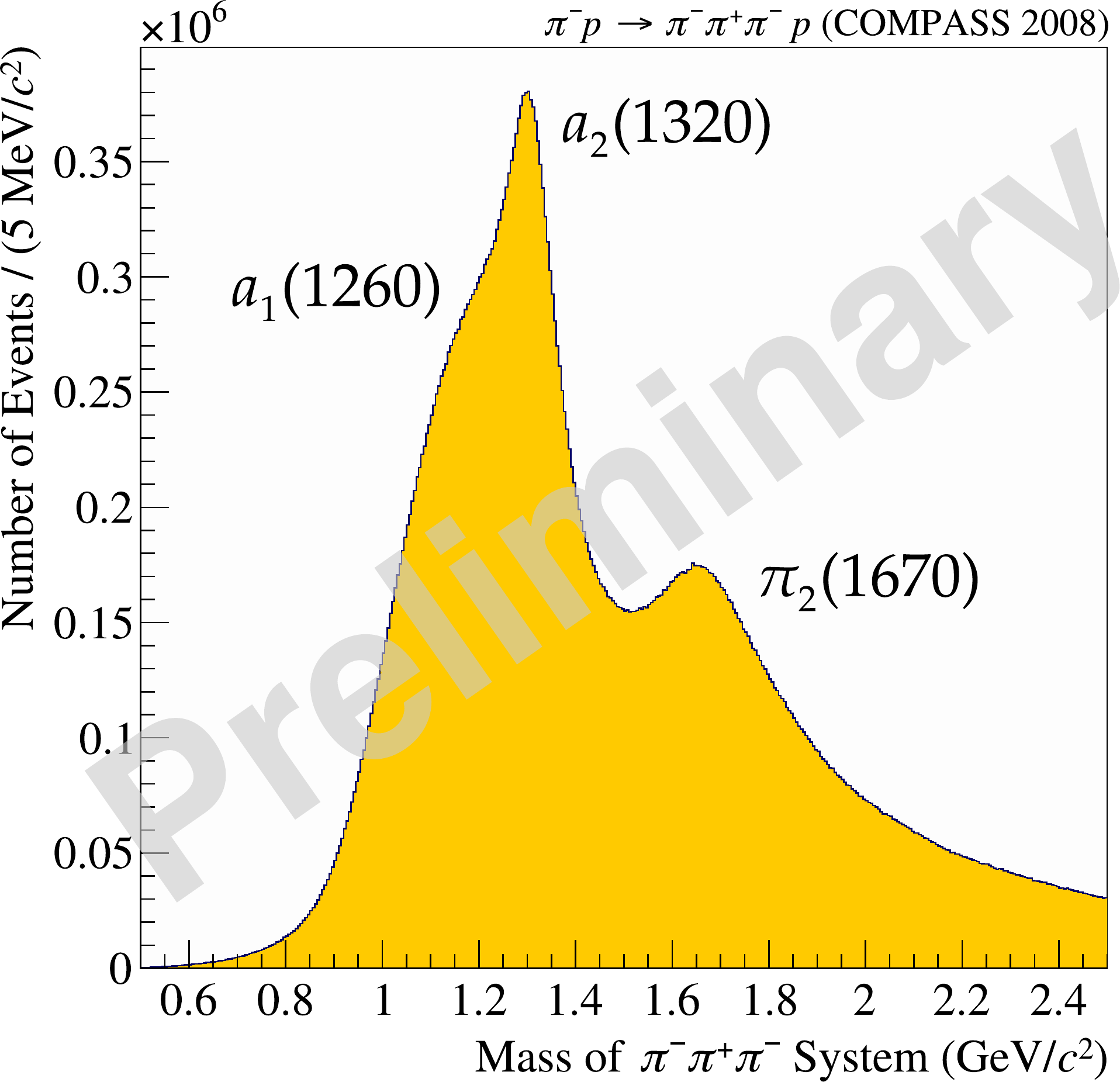}
  \includegraphics[width=0.456\textwidth]{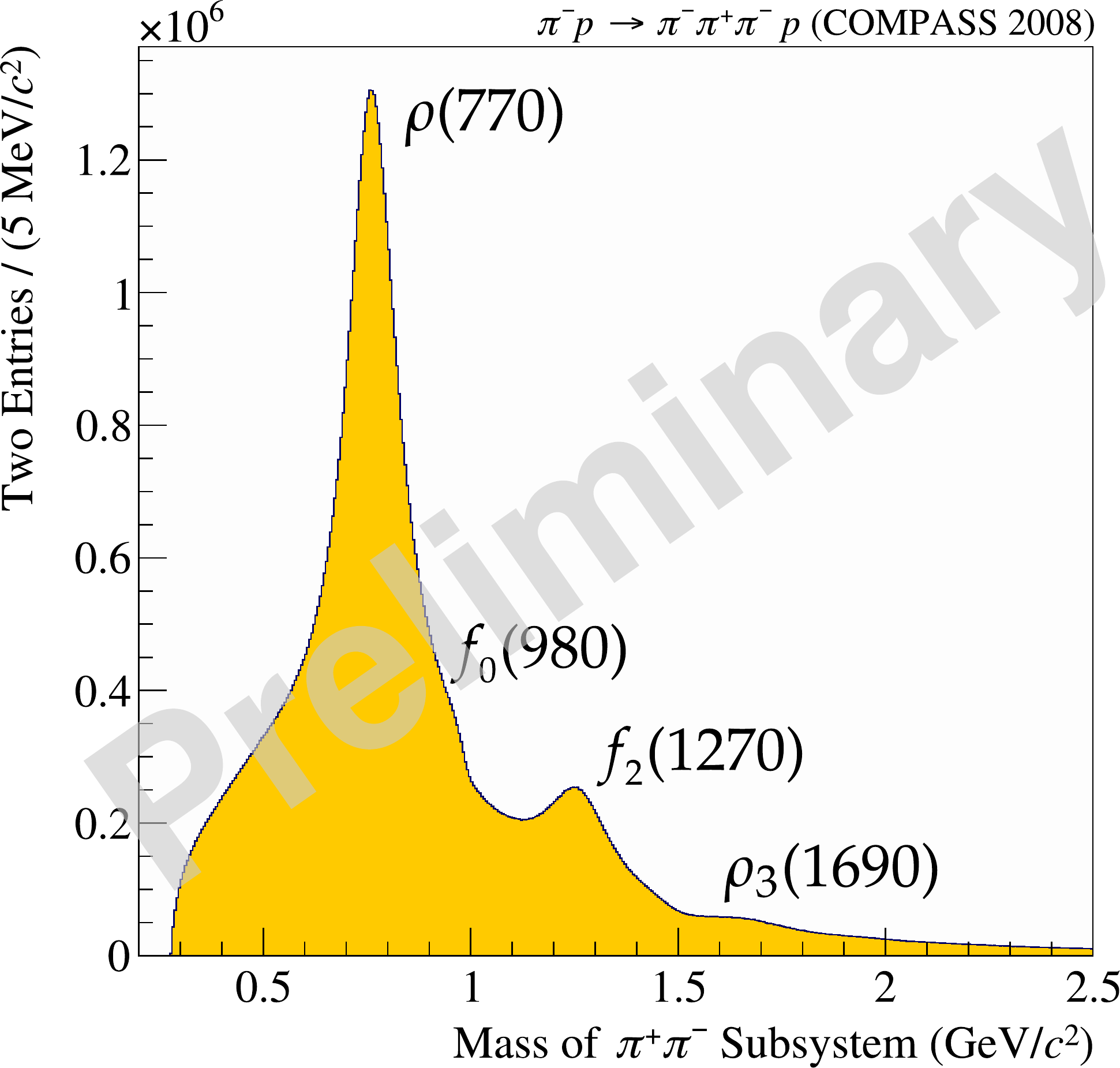}
  \caption{Left: $\pi^-\pi^+\pi^-$ invariant mass spectrum in the
    analyzed range; Right: $\pi^+\pi^-$ mass distribution.}
  \label{fig:mass}
\end{figure}

The COMPASS experiment has recorded a large data set of the
diffractive dissociation reaction
$\pi^- + p \to (3\pi)^- + p_\text{recoil}$ using a 190~GeV/$c$ pion
beam on a liquid-hydrogen target.  This reaction is known to exhibit a
rich spectrum of produced intermediate three-pion states.  In the
past, several candidates for spin-exotic mesons have been reported in
pion-induced diffraction~\cite{exotic}.  In diffractive reactions the
beam hadron is excited to some intermediate state $X^-$ via
$t$-channel Reggeon exchange with the target.  At 190~GeV/$c$ beam
momentum, pomeron exchange is dominant.  In the reaction considered
here, the $X$ decays into the $\pi^-\pi^+\pi^-$ and $\pi^-\pi^0\pi^0$
final states, which are detected by the spectrometer.  The scattering
process is characterized by two kinematic variables: the squared total
center-of-mass energy s and the squared four-momentum transfer to the
target $t = (p_\text{beam} - p_{X})^2 < 0$.  It is customary to use
the reduced four-momentum transfer $t' \equiv |t| - |t|_\text{min}$
instead of $t$, where $|t|_\text{min}$ is the minimum value of $|t|$
for a given invariant mass of $X$.  After all cuts, the data sample
consists of $46 \times 10^6$ $\pi^-\pi^+\pi^-$ and $3.5 \times 10^6$
$\pi^-\pi^0\pi^0$ exclusive events in the analyzed kinematic region of
three-pion mass, $0.5 < m_{3\pi} < 2.5~\text{GeV}/c^2$, and
four-momentum transfer squared, $0.1 < t' < 1.0~(\text{GeV}/c)^2$.
Figure~\ref{fig:mass} shows the $\pi^-\pi^+\pi^-$ mass spectrum as
well as that of the $\pi^+\pi^-$ subsystem.  The known pattern of
resonances $a_1(1260)$, $a_2(1320)$, and $\pi_2(1670)$ is seen in the
$3\pi$ system along with $\rho(770)$, $f_0(980)$, $f_2(1270)$, and
$\rho_3(1690)$ in the $\pi^+\pi^-$ subsystem.

\section{Partial-Wave Decomposition}

In order to disentangle the different contributing intermediate $3\pi$
states $X$, a partial-wave analysis (PWA) was performed.  The PWA is
based on the isobar model, which assumes that the $X$ decays first
into an intermediate resonance, which is called the isobar, and a
``bachelor pion''.  In a second step, the isobar decays into
$\pi^+\pi^-$.  In accordance with the $\pi^+\pi^-$ invariant mass
spectrum shown in Fig.~\ref{fig:mass} and with analyses by previous
experiments, we include $[\pi\pi]_S$, $\rho(770)$, $f_0(980)$,
$f_2(1270)$, $f_0(1500)$, and $\rho_3(1690)$ as isobars into the fit
model.  Here, $[\pi\pi]_S$ represents the broad component of the
$\pi\pi$ $S$-wave.  Based on the six isobars, we have constructed a
set of partial waves that consists of 88 waves in total, including one
non-interfering flat wave representing three uncorrelated pions.  This
constitues the largest wave set ever used in an analysis of the $3\pi$
final state.  The partial-wave decomposition is performed in narrow
bins of the $3\pi$ invariant mass and makes no assumptions on the
$3\pi$ resonance content of the partial waves.  Each $m_{3\pi}$ bin is
further subdivided into non-equidistant bins in the four-momentum
transfer $t'$.  For the $\pi^-\pi^+\pi^-$ channel 11 bins are used,
for the $\pi^-\pi^0\pi^0$ final state 8 bins.  With this additional
binning in $t'$, the dependence of the partial-wave amplitudes on the
four-momentum transfer can be studied in detail.  The details of the
analysis model are described in Ref.~\cite{long_paper}.

The partial-wave amplitudes are extracted from the data as a function
of $m_{3\pi}$ and $t'$ by fitting the five-dimensional kinematic
distributions of the outgoing three pions.  The amplitudes do not only
contain information about the partial-wave intensities, but also about
the relative phases of the partial waves.  The latter are crucial for
resonance extraction.

Partial waves are defined by the quantum numbers of the $X$ (spin $J$,
parity $P$, $C$-parity, spin projection $M$), the naturality
$\epsilon = \pm 1$ of the exchange particle, the isobar, and the
orbital angular momentum $L$ between the isobar and the bachelor pion.
These quantities are summarized in the partial-wave notation
$J^{PC}\,M^\epsilon\,\text{[isobar]}\,\pi\,L$.  Since at the used beam
energies pomeron exchange is dominant, 80 of the 88 partial waves in
the model have $\epsilon = +1$.  The $C$-parity is by convention that
of the neutral isospin partner of the $X^-$.

\section{The $J^{PC} = 1^{-+}$ Spin-Exotic Wave}

\begin{figure}[tb]
  \centering
  \includegraphics[width=0.45\textwidth]{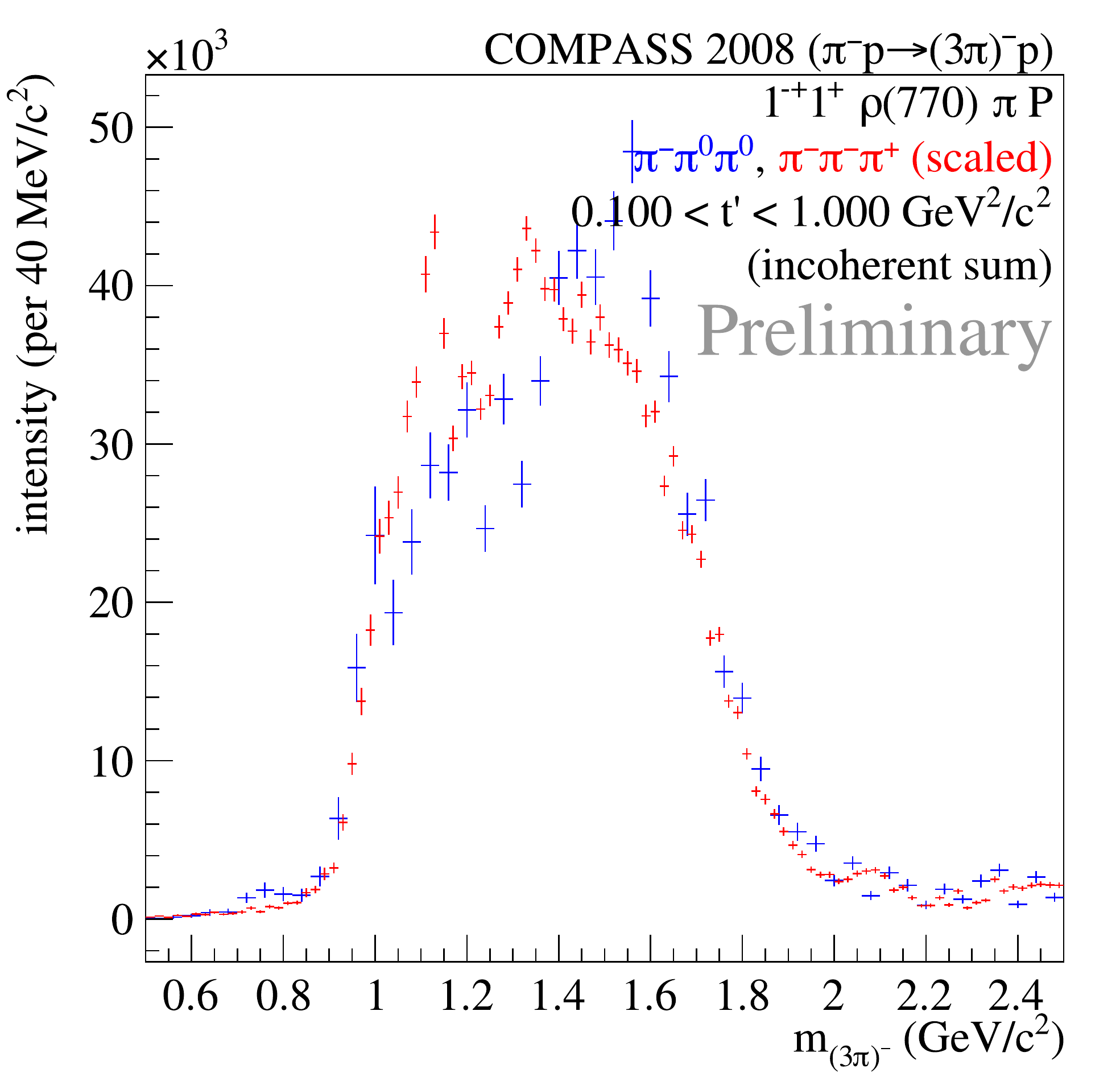}
  \caption{Intensity of the $1^{-+}\,1^+\,\rho(770)\,\pi\,P$ wave
    summed over all $t'$ bins for the $\pi^-\pi^+\pi^-$ (red) and the
    $\pi^-\pi^0\pi^0$ (blue) final state.}
  \label{fig:1mp_intens}
\end{figure}

The 88-wave model contains also spin-exotic waves with $J^{PC}$
quantum numbers that are forbidden for quark-antiquark states in the
non-relativistic limit.  The most interesting of these waves is the
$1^{-+}\,1^+\,\rho(770)\,\pi\,P$ wave, which contributes less than 1\%
to the total intensity.  Previous analyses claimed a resonance, the
$\pi_1(1600)$, at about 1.6~GeV/$c^2$ in this
channel~\cite{bnl_1,compass_pb}.  Figure~\ref{fig:1mp_intens} shows
the intensity of this partial wave for the two final states
($\pi^-\pi^+\pi^-$ in red, $\pi^-\pi^0\pi^0$ in blue).  The two
distributions are scaled to have the same integral.

\begin{figure}[tb]
  \centering
  \includegraphics[width=0.45\textwidth]{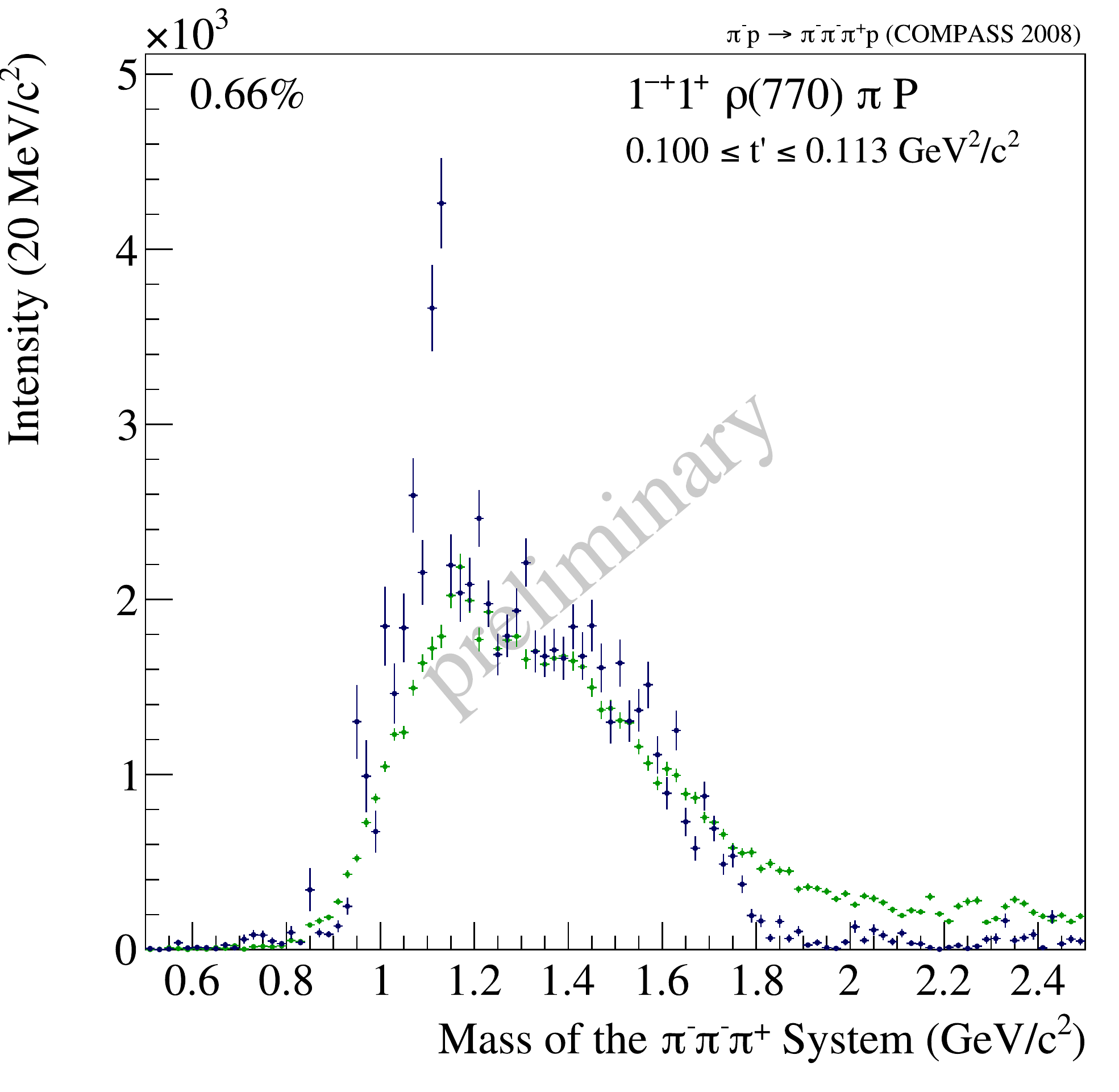}
  \includegraphics[width=0.45\textwidth]{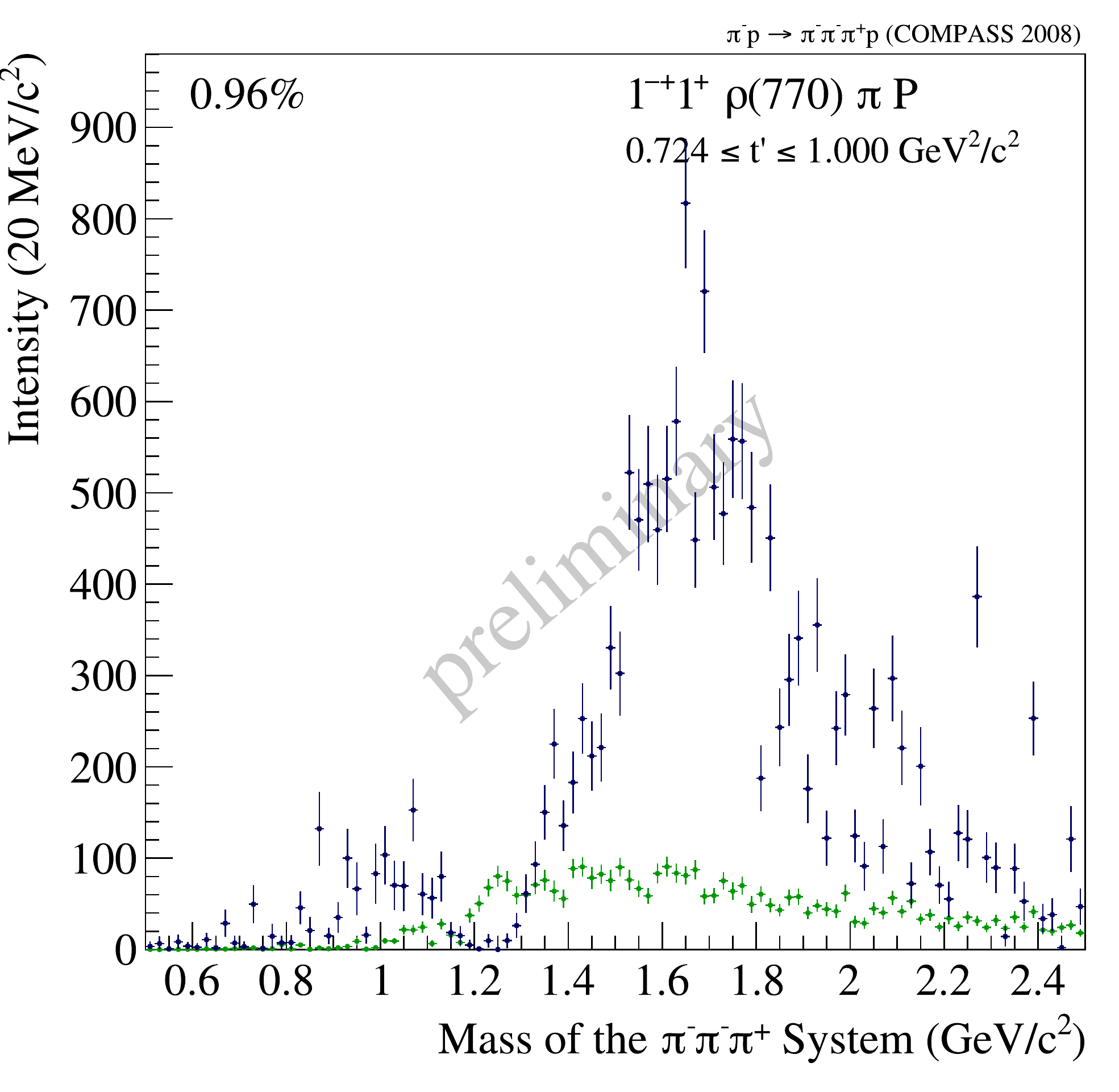}
  \caption{Intensity of the $1^{-+}\,1^+\,\rho(770)\,\pi\,P$ wave in
    different regions of $t'$ for the $\pi^-\pi^+\pi^-$ final state
    (dark blue).  The partial-wave projections of Monte-Carlo data
    generated according to a model of the Deck effect are overlaid in
    green.}
  \label{fig:1mp_tbins}
\end{figure}

\begin{figure}[tb]
  \centering
  \includegraphics[width=0.45\textwidth]{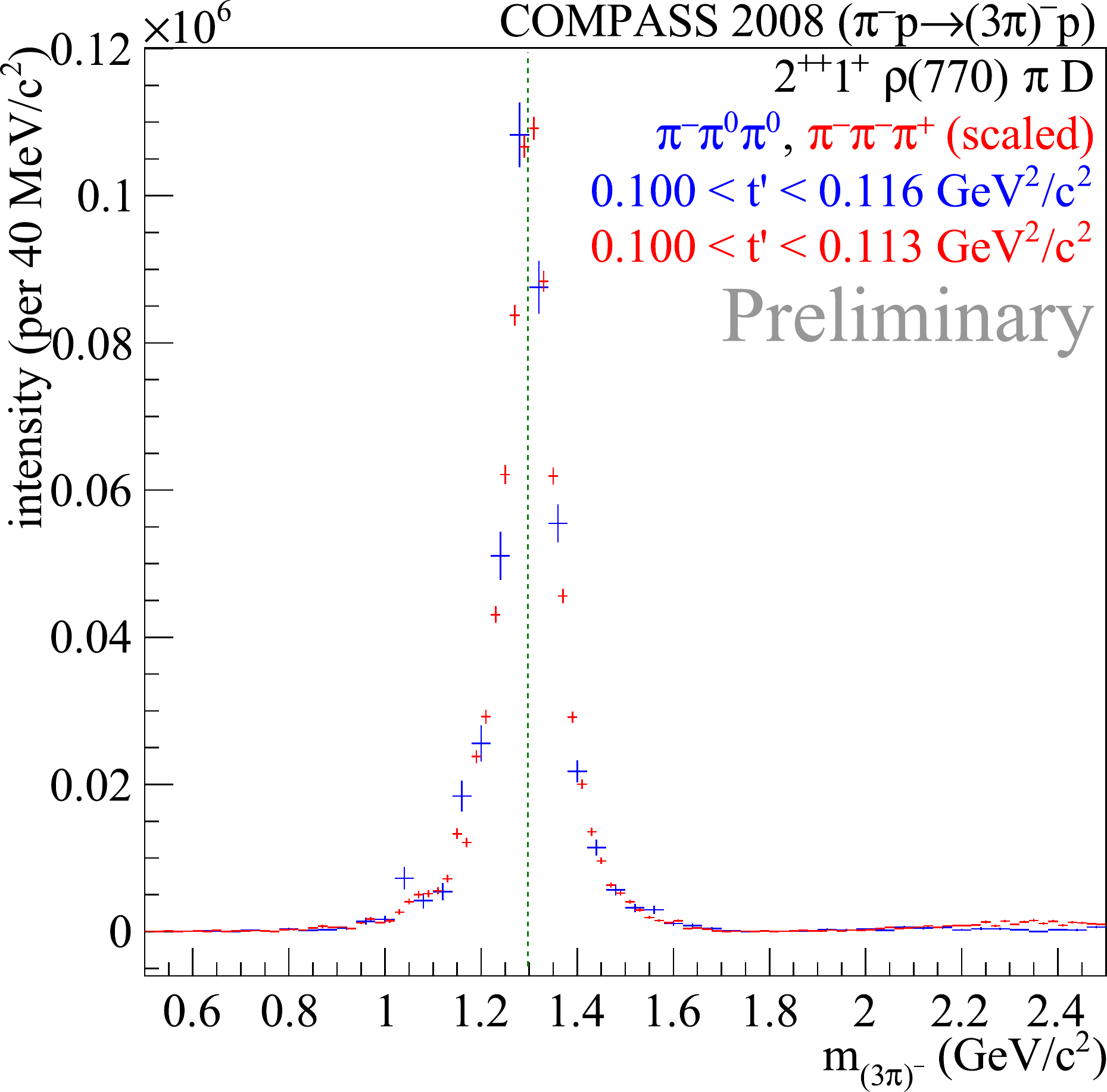}
  \includegraphics[width=0.45\textwidth]{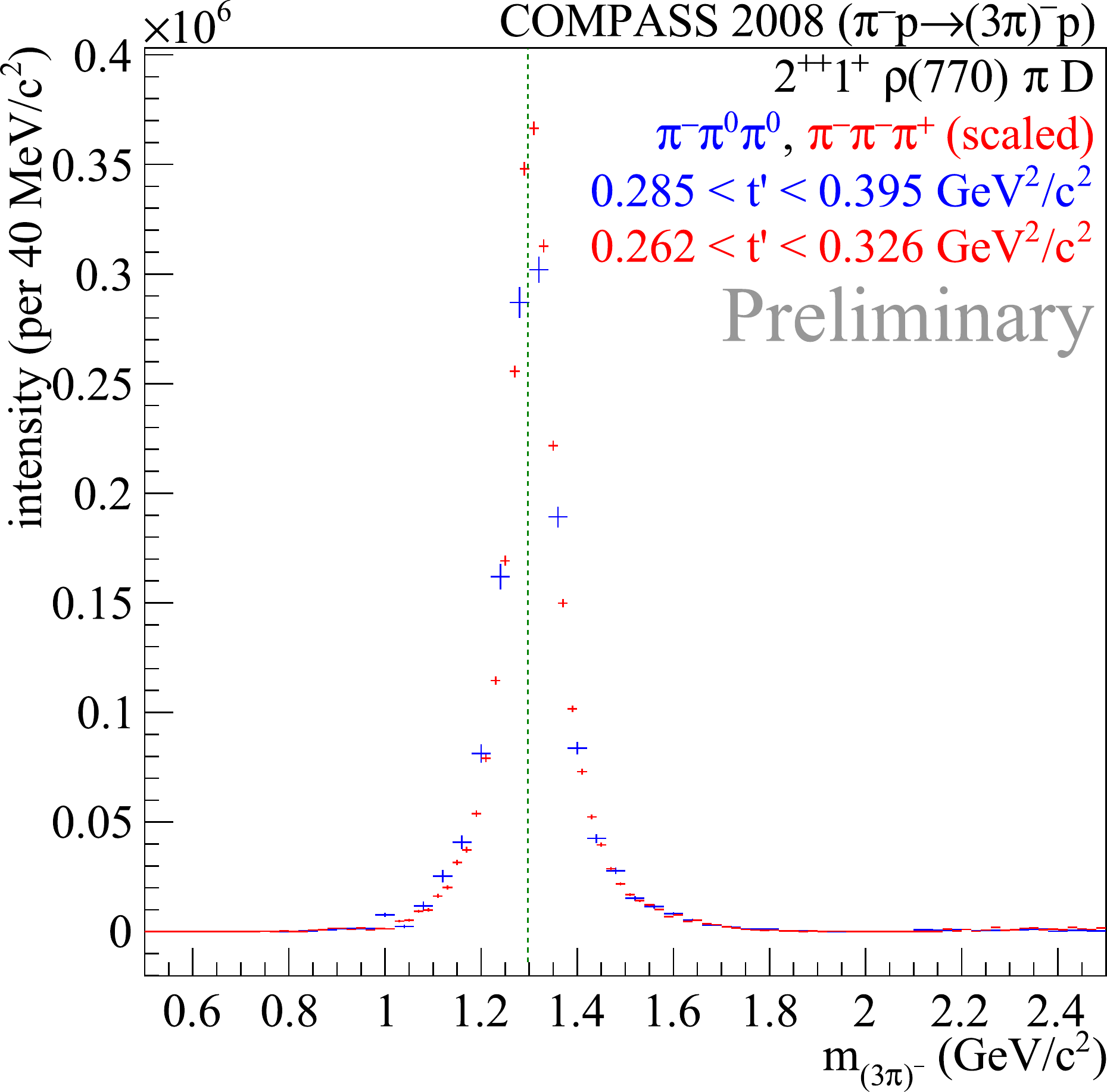}
  \caption{Intensity of the $2^{++}\,1^+\,\rho(770)\,\pi\,D$ wave in
    different regions of $t'$ for the $\pi^-\pi^+\pi^-$ (red) and the
    $\pi^-\pi^0\pi^0$ (blue) final state.}
  \label{fig:2pp_tbins}
\end{figure}

Both decay channels are in fair agreement and exhibit a broad
enhancement extending from about 1.0 to 1.8~GeV/$c^2$.  In the 1.0 to
1.2~GeV/$c^2$ mass range the intensity depends strongly on the details
of the fit model.  Peak-like structures in this region are probably
due to imperfections of the applied PWA model.  A remarkable change of
the shape of the intensity spectrum of the
$1^{-+}\,1^+\,\rho(770)\,\pi\,P$ wave with $t'$ is observed (see
Fig.~\ref{fig:1mp_tbins}).  At values of $t'$ below about
0.3~$(\text{GeV}/c)^2$, we observe no indication of a resonance peak
around $m_{3\pi} = 1.6$~GeV/$c^2$, where we would expect the
$\pi_1(1600)$.  For the $t'$ bins in the interval from 0.449 to
$1.000~(\text{GeV}/c)^2$, the observed intensities exhibit a very
different shape as compared to the low-$t'$ region, with a structure
emerging at about 1.6~GeV/$c^2$ and the intensity at lower masses
disappearing rapidly with increasing $t'$.  This is in contrast to
clean resonance signals like the $a_2(1320)$ in the
$2^{++}\,1^+\,\rho(770)\,\pi\,D$ wave, which, as expected, do not
change their shape with $t'$ (see Fig.~\ref{fig:2pp_tbins}).  The
observed $t'$ behavior of the $1^{-+}$ wave is therefore a strong
indication that non-resonant contributions play a dominant role.

It is believed that the non-resonant contribution in the $1^{-+}$ wave
originates predominantly from the Deck effect, in which the incoming
beam pion dissociates into the isobar and an off-shell pion that
scatters off the target proton to become on-shell~\cite{deck}.  As a
first step towards a better understanding of the non-resonant
contribution, Monte-Carlo data were generated that are distributed
according to a model of the Deck effect.  The model employed here is
very similar to that used in Ref.~\cite{accmor_deck}.  The
partial-wave projection of these Monte Carlo data is shown as green
points in Fig.~\ref{fig:1mp_tbins}.  In order to compare the
intensities of real data and the Deck-model pseudo data, the Monte
Carlo data are scaled to the $t'$-summed intensity of the $1^{-+}$
wave as observed in real data.  At values of $t'$ below about
0.3~$(\text{GeV}/c)^2$, the intensity distributions of real data and
Deck Monte Carlo exhibit strong similarities suggesting that the
observed intensity might be saturated by the Deck effect.  Starting
from $t' \approx 0.4~(\text{GeV}/c)^2$, the spectral shapes for Deck
pseudo-data and real data deviate from each other with the differences
increasing towards larger values of $t'$.  This leaves room for a
potential resonance signal.  It should be noted, however, that the
Deck pseudo data contain no resonant contributions. Therefore,
potential interference effects between the resonant and non-resonant
amplitudes cannot be assessed in this simple approach.

\section{The $a_1(1420)$}

\begin{figure}[tb]
  \centering
  \includegraphics[width=0.45\textwidth]{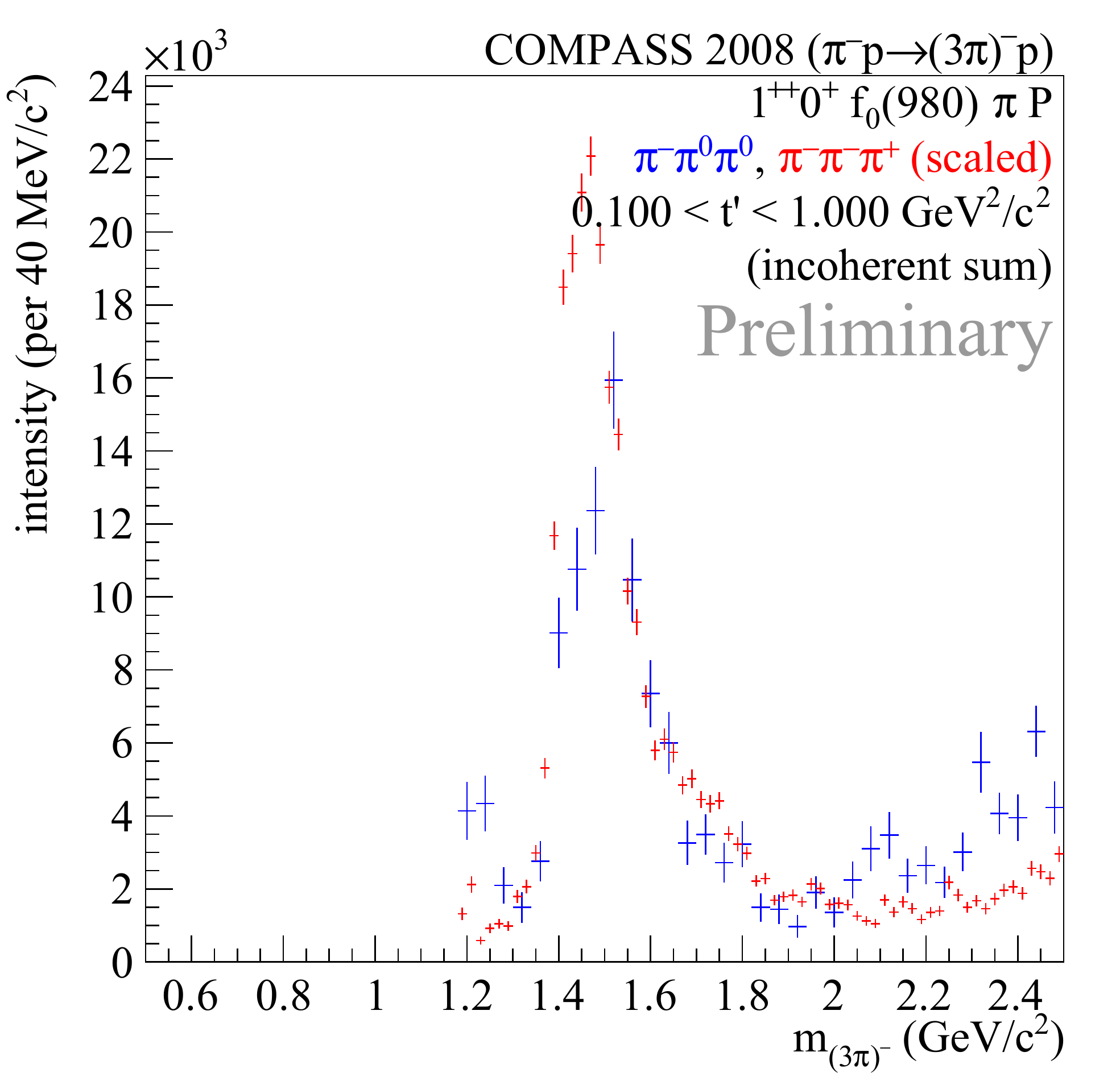}
  \includegraphics[width=0.45\textwidth]{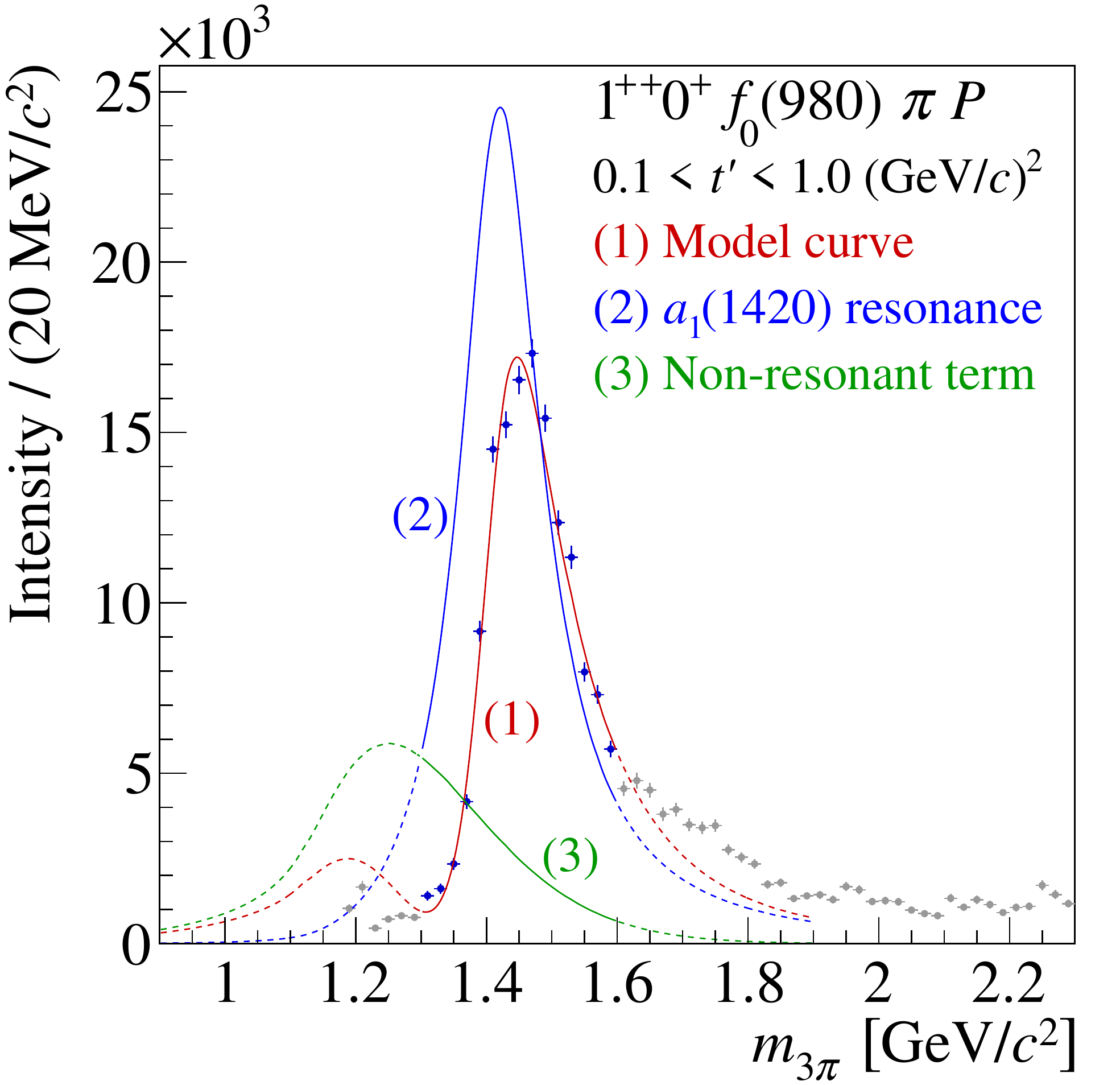}
  \caption{Left: Intensity of the $1^{++}\,0^+\,f_0(980)\,\pi\,P$ wave
    summed over all $t'$ bins for the $\pi^-\pi^+\pi^-$ (red) and the
    $\pi^-\pi^0\pi^0$ (blue) final states.  Right: Result of a
    resonance-model fit to the $\pi^-\pi^+\pi^-$ data~\cite{a1_1420}.
    The data points correspond to the red points in the left figure.}
  \label{fig:1pp_f0_intens}
\end{figure}

\begin{figure}[tb]
  \centering
  \includegraphics[width=0.45\textwidth]{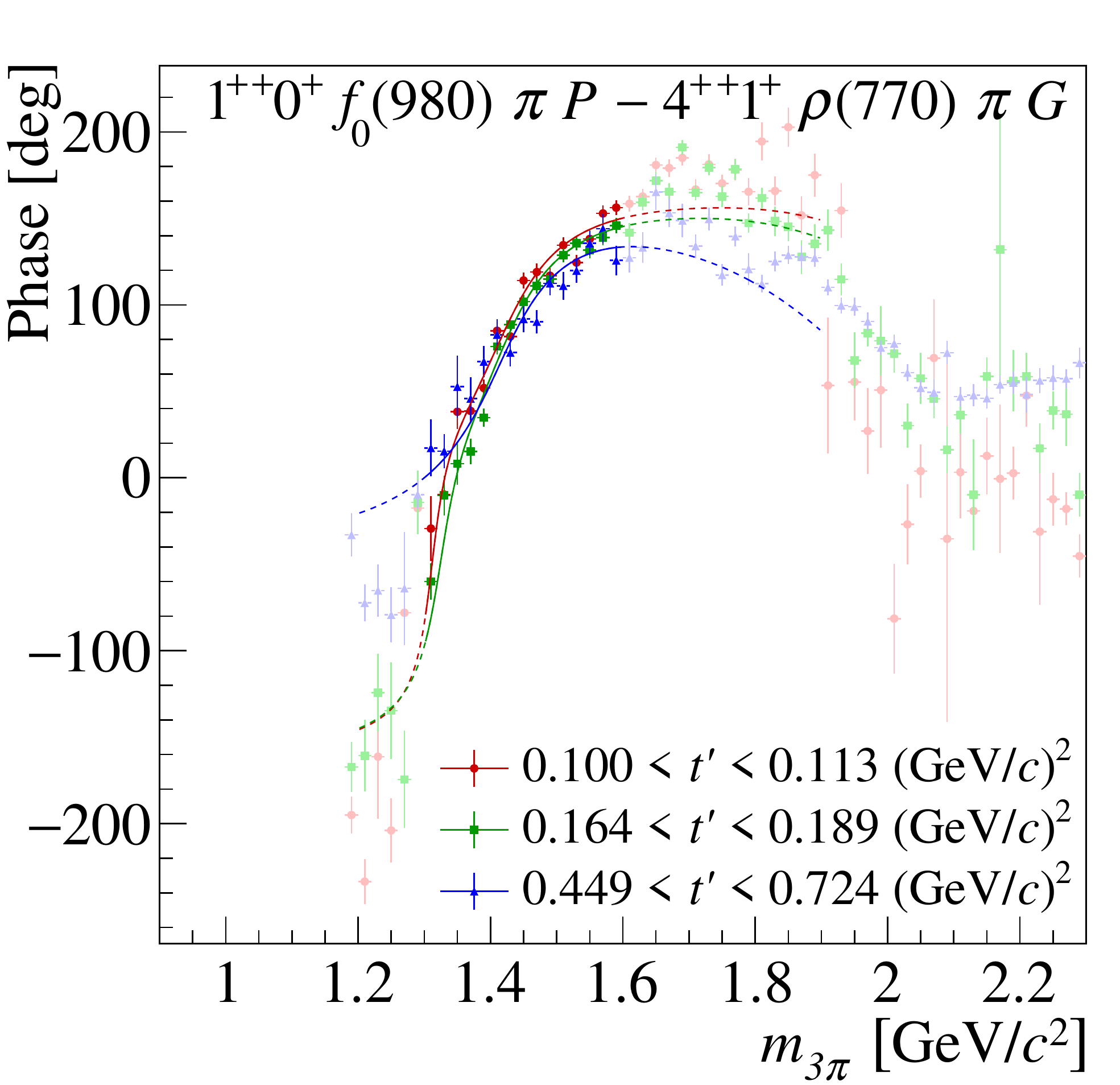}
  \includegraphics[width=0.45\textwidth]{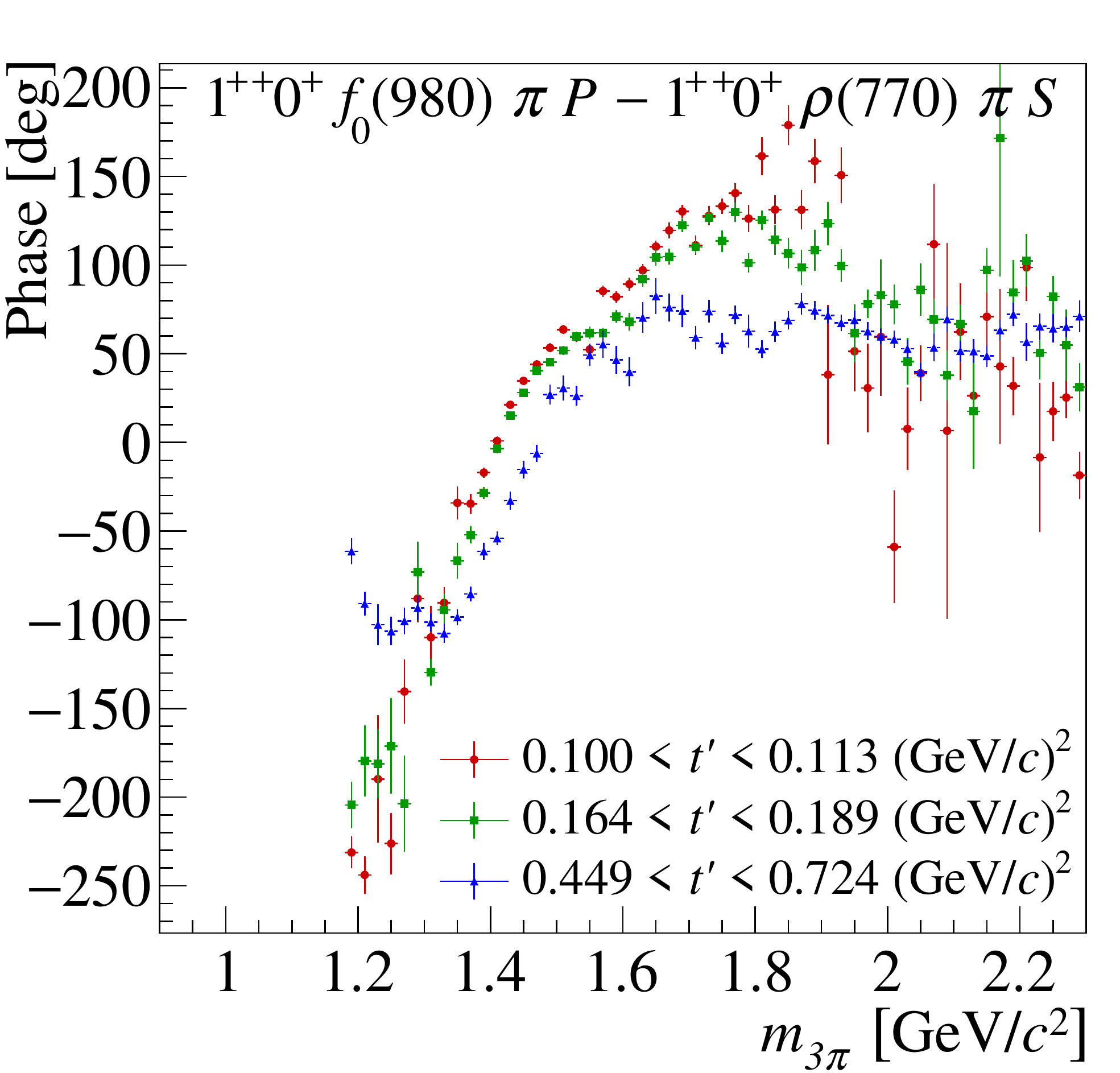}
  \caption{Examples for relative phases of the
    $1^{++}\,0^+\,f_0(980)\,\pi\,P$ wave with respect to the
    $4^{++}\,1^+\,\rho(770)\,\pi\,G$ (left) and the
    $1^{++}\,0^+\,\rho(980)\,\pi\,S$ wave (right)~\cite{a1_1420}.  The
    phases are shown for three different $t'$ regions indicated by the
    color.}
  \label{fig:1pp_f0_phase}
\end{figure}

A surprising find in the COMPASS data was a pronounced narrow peak at
about 1.4~GeV/$c^2$ in the $1^{++}\,0^+\,f_0(980)\,\pi\,P$ wave (see
Fig.~\ref{fig:1pp_f0_intens}).  The peak is observed with similar
shape in the $\pi^-\pi^+\pi^-$ and $\pi^-\pi^0\pi^0$ data and is
robust against variations of the PWA model.  In addition to the peak
in the partial-wave intensity, rapid phase variations with respect to
most waves are observed in the 1.4~GeV/$c^2$ region (see
Fig.~\ref{fig:1pp_f0_phase}).  The phase motion as well as the peak
shape change only little with $t'$.

In order to test the compatibility of the signal with a Breit-Wigner
resonance, a resonance-model fit was performed using a novel method,
where the intensities and relative phases of three waves
($1^{++}\,0^+\,f_0(980)\,\pi\,P$, $2^{++}\,1^+\,\rho(770)\,\pi\,D$,
and $4^{++}\,1^+\,\rho(770)\,\pi\,G$) were fit simultaneously in all
11 $t'$ bins~\cite{a1_1420}.  Forcing the resonance parameters to be
the same across all $t'$ bins leads to an improved separation of
resonant and non-resonant contribution as compared to previous
analyses that did not incorporate the $t'$ information.  The
Breit-Wigner model describes the peak in the
$1^{++}\,0^+\,f_0(980)\,\pi\,P$ wave well and yields a mass of
$m_0 = 1414^{+15}_{-13}$~MeV/$c^2$ and a width of
$\Gamma_0 = (153^{+8}_{-23})$~MeV/$c^2$ for the $a_1(1420)$.  Due to
the high precision of the data, the uncertainties are dominated by
systematic effects.

The $a_1(1420)$ signal is remarkable in many ways.  It appears in a
mass region that is well studied since decades.  However, previous
experiments were unable to see the peak, because it contributes only
0.25\% to the total intensity.  The $a_1(1420)$ is very close in mass
to the $1^{++}$ ground state, the $a_1(1260)$.  But it has a much
smaller width than the $a_1(1260)$.  The $a_1(1420)$ peak is seen only
in the $f_0(980)\,\pi$ decay mode of the $1^{++}$ waves and lies
suspiciously close to the $K\, \bar{K}^*(892)$ threshold.

The nature of the $a_1(1420)$ is still unclear and several
interpretations were proposed.  It could be the isospin partner to the
$f_1(1420)$.  It was also described as a two-quark-tetraquark mixed
state~\cite{wang} and a tetraquark with mixed flavor
symmetry~\cite{chen}.  Other models do not require an additional
resonance:
Ref.~\cite{berger1,berger2} proposes resonant re-scattering
corrections in the Deck process as an explanation, whereas
Ref.~\cite{bonn} suggests a branching point in the triangle
rescattering diagram for
$a_1(1260) \to K\, \bar{K}^*(892) \to K\, \bar{K}\, \pi \to f_0(980)\,
\pi$.
More detailed studies are needed in order to distinguish between these
models.

\section*{Acknowledgments}
This work was supported by the BMBF, the MLL and the Cluster of
Excellence Exc153 ``Origin and Structure of the Universe''.

\end{document}